\documentclass[fleqn,usenatbib]{mnras}

\usepackage{newtxtext,newtxmath}
\usepackage[T1]{fontenc}
\usepackage{ae,aecompl}

\usepackage{graphicx}	% Including figure files
\usepackage{amsmath}	% Advanced maths commands
\usepackage{amssymb}	% Extra maths symbols
\usepackage{siunitx,url}

\title[How to detect the shortest period binary pulsars]{How to detect
the shortest period binary pulsars in the era of \textit{LISA}}

\author[K. Kyutoku, Y. Nishino and N. Seto]{Koutarou
Kyutoku$^{1,2,3,4}$\thanks{E-mail: kyutoku@post.kek.jp}, Yuki
Nishino$^5$, and Naoki Seto$^5$\\
$^1$Theory Center, Institute of Particle and Nuclear Studies, KEK,
Tsukuba 305-0801, Japan\\
$^2$Department of Particle and Nuclear Physics, the Graduate University
for Advanced Studies (Sokendai), Tsukuba 305-0801, Japan\\
$^3$Interdisciplinary Theoretical and Mathematical Sciences Program
(iTHEMS), RIKEN, Wako, Saitama 351-0198, Japan\\
$^4$Center for Gravitational Physics, Yukawa Institute for Theoretical
Physics, Kyoto University, Kyoto 606-8502, Japan\\
$^5$Department of Physics, Kyoto University, Kyoto 606-8502, Japan
}

\date{Accepted XXX. Received YYY; in original form ZZZ}

\pubyear{2018}

\begin{document}
\label{firstpage}
\pagerange{\pageref{firstpage}--\pageref{lastpage}}
\maketitle

\begin{abstract}
 We discuss a multimessenger strategy to detect radio pulses from
 Galactic binary neutron stars in a very tight orbit with the period
 shorter than 10 min. On one hand, all-sky surveys by radio instruments
 are inefficient for detecting faint pulsars in very tight binaries due
 partly to the rarity of targets and primarily to the need of correction
 for severe Doppler smearing. On the other hand, the \textit{Laser
 Interferometer Space Antenna (LISA)} will detect these binaries with a
 very large signal-to-noise ratio and determine the orbital frequency,
 binary parameters, and sky location to high accuracy. The information
 provided by \textit{LISA} will reduce the number of required pointings
 by two to six orders of magnitude and that of required trials for the
 corrections by about nine orders of magnitude, increasing the chance of
 discovering radio pulsars. For making full use of this strategy, it is
 desirable to operate high-sensitivity radio instruments such as Square
 Kilometer Array Phase 2 simultaneously with \textit{LISA}.
\end{abstract}


%%%%%%%%%%%%%%%%%%%%%%%%%%%%%%%%%%%%%%%%%%%%%%%%%%

%%%%%%%%%%%%%%%%% BODY OF PAPER %%%%%%%%%%%%%%%%%%

\section{Introduction} \label{sec:intro}

Radio pulsars are fascinating laboratories of astrophysics
\citep{lorimer_kramer}. In particular, those in short-period binaries
serve as useful tools to study binary evolution \citep{tauris_etal2017},
interacting magnetospheres \citep[e.g.][]{lyne_etal2004}, and the theory
of gravitation \citep{will2014}. Even after direct detections of
gravitational waves from compact binary coalescences, human time-scale
monitoring of relativistic binary pulsars could remain undefeatable for
constraining some models of modified gravity
\citep{yunes_yp2016,ligovirgo2018-4}. Furthermore, these binaries could
be our Schelling point to communicate with extraterrestrial intelligence
\citep{nishino_seto2018}.

The tightest Galactic binary neutron stars should have much shorter
orbital periods than those observed today. On one hand, as of 2018, the
shortest period of binary neutron stars hosting a detectable pulsar is
\SI{1.88}{h} of J1946$+$2052 \citep{stovall_etal2018}, which will spend
\SI{46}{Myr} before merger. On the other hand, the gravitational-wave
event GW170817 suggests that the Galactic merger rate of binary neutron
stars is about one in \SI{e4}{yr} \citep{ligovirgo2017-3}. Therefore, it
is virtually certain that many binary neutron stars with short orbital
periods and being close to merger exist in our Galaxy. It is also
natural to expect that some of them host detectable radio pulsars.

Pulsars in very tight binaries, however, are difficult to find, because
strong Doppler smearing has to be corrected appropriately. While the
acceleration search has been adopted to discover binary pulsars
\citep{faulkner_etal2004}, this method assumes a linear drift of the
frequency and is applicable only to integration times shorter than
$\approx \SI{10}{per~cent}$ of the orbital period
\citep{johnston_kulkarni1991}. Thus, pulsars in very tight binaries
easily elude this search unless they are extremely luminous. Jerk
searches are proposed to improve the situation, but integration times
are still restricted to $\lesssim \SI{15}{per~cent}$ of the orbital
period \citep{bagchi_lw2013,andersen_ransom2018}. The incoherent
phase-modulation search is applicable to integration times longer than
the orbital period \citep{jouteux_rsjv2002,ransom_ce2003}.

In this paper, we examine an alternative and coherent method to detect
pulsars in very tight binaries utilizing information provided by a
future space-borne gravitational-wave detector, the \textit{Laser
Interferometer Space Antenna} \citep[\textit{LISA}; see][for the results
of \textit{LISA} Pathfinder]{lisapf2016,lisapf2018}. \textit{LISA} is
sensitive at \si{\milli\hertz} bands, and its targets including various
compact object binaries are summarized in \citet{lisa2017}. As we will
show later, the merger rate estimated by GW170817 suggests that a number
of Galactic binary neutron stars can also be detected with \textit{LISA}
as quasi-monochromatic sources, where $O(10)$ of them are free from the
foreground associated with white dwarf binaries. \textit{LISA} will
provide us with accurate information of these binaries such as the
orbital frequency (or period), binary parameters, and sky location
\citep{cutler1998,seto2002,takahashi_seto2002}. This information enables
pulsar surveys by radio instruments to reduce both the searching time
and the computational cost of corrections for Doppler smearing
significantly. Although we only focus on binary neutron stars for
concreteness, our discussion also applies to black hole/white
dwarf--neutron star binaries.

This paper is organized as follows. We first review the prospect for
detecting Galactic binary neutron stars with \textit{LISA} in Section
\ref{sec:gw}. Next, we discuss how to detect radio pulses from these
binaries in Section \ref{sec:radio}, comparing performance of our
multimessenger strategy with that of an all-sky survey solely by radio
instruments in Section \ref{sec:radio_comp}. Section \ref{sec:summary}
is devoted to a summary. In this paper, $G$, $c$, and $k_\mathrm{B}$
denote the gravitational constant, speed of light, and Boltzmann's
constant, respectively.

\section{Gravitational-wave observation by \textit{LISA}} \label{sec:gw}

First, we estimate statistical errors in extracting parameters of
Galactic binary neutron stars with \textit{LISA}. Gravitational-wave
frequency is denoted by $f$ and the operation period of \textit{LISA} is
denoted by $t_\mathrm{LISA}$. We adopt the quadrupole approximation in
this work, and the gravitational-wave frequency is twice the orbital
frequency except for higher harmonics associated with the eccentricity.

\subsection{Target}

The high rate of binary neutron star mergers reported by the LIGO--Virgo
collaboration, $R = 1540^{+3200}_{-1220} \, \si{Gpc^{-3}.yr^{-1}}$
\citep{ligovirgo2017-3}, suggests that Galactic binary neutron stars are
numerous. The Galactic merger rate may be given by $R_\mathrm{MW}
\approx R / n_\mathrm{MW}$, where $n_\mathrm{MW} \approx
\SI{0.01}{Mpc^{-3}}$ is the number density of Milky Way equivalent
galaxies. Assuming that the Galactic population of binary neutron stars
is stationary on the time-scale of interest, their distribution in
frequency is given by \citep[e.g.][]{kyutoku_seto2016}
\begin{equation}
 \frac{dN_\mathrm{MW}}{df} = \dot{f}^{-1} R_\mathrm{MW} .
\end{equation}
The time derivative of gravitational-wave frequency, or the chirp
parameter, is given by \citep{peters1964}
\begin{align}
 \dot{f} & = \frac{96 \pi^{8/3} ( G \mathcal{M} )^{5/3} f^{11/3}}{5
 c^5} \\
 & = \SI{1.3e-15}{\hertz\per\second} \left(
 \frac{f}{\SI{4}{\milli\hertz}} \right)^{11/3} \left(
 \frac{\mathcal{M}}{1.2 M_\odot} \right)^{5/3} \label{eq:chirp},
\end{align}
where $\mathcal{M}$ is the chirp mass of the binary, in the quadrupole
approximation. Hence, the number of Galactic binary neutron stars
emitting gravitational waves at frequency higher than $f$ is estimated
to be
\begin{align}
 N_\mathrm{MW}(>f) & = \int_f^\infty \frac{dN_\mathrm{NW}}{df'} df' \\
 & = \frac{5 c^5 R_\mathrm{MW}}{256 \pi^{8/3} ( G \mathcal{M} )^{5/3}
 f^{8/3}} \\
 & = 5.6 \left( \frac{\mathcal{M}}{1.2 M_\odot} \right)^{-5/3} \left(
 \frac{f}{\SI{4}{\milli\hertz}} \right)^{-8/3} \notag \\
 & \times \left( \frac{R_\mathrm{MW}}{\SI{1.5e-4}{yr^{-1}}} \right) .
\end{align}

This estimate indicates that the shortest period class of Galactic
binary neutron stars will be observed via gravitational waves in mHz
bands by \textit{LISA} \citep{lisa2017}. The frequency evolution during
the operation period of \textit{LISA} is insignificant for binaries with
$f \ll \SI{100}{\milli\hertz}$ considered here, and thus we may regard
them as quasi-monochromatic. Although numerous binary neutron stars
exist also at $f \lesssim \SI{3}{\milli\hertz}$, which corresponds to
the orbital period longer than \SI{10}{\min}, the sensitivity of
\textit{LISA} will be degraded because of the foreground associated with
unresolved white dwarf binaries \citep{nelemans_yp2004,robson_cl2018}.

Galactic binary neutron stars may be classified into two classes
according to their eccentricities \citep[see][for a
compilation]{tauris_etal2017}. If we focus only on binaries that merge
within the Hubble time, more than the half will have low eccentricities
of $e < \num{5e-3}$ at $f = \SI{4}{\milli\hertz}$ due to gravitational
radiation reaction \citep{peters1964} and may be regarded as circular
for our purpose. However, three binaries, namely PSR B1913$+$16
\citep{hulse_taylor1975,weisberg_huang2016}, B2127$+$11C
\citep{prince_akw1991,jacoby_cjamk2006}, and J1757$-$1854
\citep{cameron_etal2018}, will retain moderate eccentricities of 0.020,
0.028, and 0.035, respectively, at \SI{4}{\milli\hertz}. In this study,
we mainly focus on circular binaries, and possible effects of moderate
eccentricity $e \approx 0.03$ are briefly discussed when they are
relevant.

\subsection{\textit{LISA} observation}

The signal-to-noise ratio $\rho$ of gravitational waves can be evaluated
by a quasi-monochromatic approximation
\citep[e.g.][]{kyutoku_seto2016}. Taking the average with respect to the
sky location and binary orientation, we have
\begin{align}
 \rho & = \frac{4 \sqrt{3} \pi^{2/3} ( G \mathcal{M} )^{5/3}}{5 c^4 D}
 \frac{f^{2/3} t_\mathrm{LISA}^{1/2}}{[(3/20) S_n (f)]^{1/2}} \\
 & = 200 \left( \frac{\mathcal{M}}{1.2M_\odot} \right)^{5/3} \left(
 \frac{D}{\SI{10}{kpc}} \right)^{-1} \left(
 \frac{t_\mathrm{LISA}}{\SI{2}{yr}} \right)^{1/2} \notag \\
 & \times \left( \frac{f}{\SI{4}{\milli\hertz}} \right)^{2/3} \left(
 \frac{S_n(f)}{\SI{2e-40}{\per\hertz}} \right)^{-1/2} , \label{eq:snr}
\end{align}
where $D$ is the distance to the binary and $S_n(f)$ is the effective
noise spectral density. We normalize $S_n(f)$ according to
\citet{robson_cl2018}. Hereafter, we use the values in this equation as
our fiducial parameters.

We estimate $1\sigma$ statistical errors of various quantities with
\textit{LISA} observations using the results derived with the Fisher
analysis in \citet{seto2002} and \citet{takahashi_seto2002}, which are
valid for $t_\mathrm{LISA} \gtrsim \SI{2}{yr}$. The extrinsic parameters
are estimated with typical errors
\begin{align}
 \frac{\Delta A}{A} & = 0.01 \left( \frac{\rho}{200} \right)^{-1} \\
 \Delta \phi & = 0.015 \left( \frac{\rho}{200} \right)^{-1}
 \label{eq:errphase} \\
 \Delta \Omega & = \SI{0.036}{deg^2} \left( \frac{\rho}{200}
 \right)^{-2} \left( \frac{f}{\SI{4}{\milli\hertz}} \right)^{-2}
 \label{eq:errsky} ,
\end{align}
where $A$, $\phi$, and $\Omega$ are the intrinsic gravitational-wave
amplitude, gravitational-wave phase, and sky location of the binary,
respectively. Because the error of the amplitude is known to be
correlated strongly with the inclination angle $\iota$, we use $\Delta
A/A$ as a proxy for the statistical error of $\Delta ( \sin \iota
)$. Although we find by repeating the Fisher analysis that this
approximation sometimes underestimates $\Delta ( \sin \iota )$ by a
factor of 2--3, it turns out later that the precise value of $\Delta (
\sin \iota )$ is not important for radio observations. Because the wave
amplitude is inversely proportional to the distance, the relative error
in the distance is also given by $\Delta A / A$ as far as the chirp mass
is determined precisely \citep{takahashi_seto2002}. Although the
Galactic electron distribution is not fully understood
\citep{yao_mw2017}, the distance error may be useful to constrain
loosely the characteristics of the target pulsar (see Section
\ref{sec:radio}).

The errors in determining the gravitational-wave frequency and its time
derivative are given by \citep{takahashi_seto2002}
\begin{align}
 \Delta f & = \SI{1.7e-10}{\hertz} \left( \frac{\rho}{200} \right)^{-1}
 \left( \frac{t_\mathrm{LISA}}{\SI{2}{yr}} \right)^{-1} \\
 \Delta \dot{f} & = \SI{5.4e-18}{\hertz\per\second} \left(
 \frac{\rho}{200} \right)^{-1} \left( \frac{t_\mathrm{LISA}}{\SI{2}{yr}}
 \right)^{-2} .
\end{align}
Note that the matched-filtering analysis, where the detector output is
cross-correlated with theoretical waveform models, allows us to estimate
these parameters more accurately than the naively estimated frequency
resolution. The gravitational-wave frequency is determined virtually
perfectly, as is the orbital frequency. Because $\Delta \mathcal{M} /
\mathcal{M} \approx (3/5) \Delta \dot{f} / \dot{f}$, where $\dot{f}$ is
given by equation \eqref{eq:chirp}, the chirp mass will be determined to
sub-per cent accuracy. Because the chirp mass of white dwarf binaries
should be small \citep{farmer_phinney2003}, binary neutron stars will be
securely selected.

Although the total mass $M$ is required to derive the orbital separation
$a = [GM/ ( \pi f )^2 ]^{1/3}$ from the gravitational-wave frequency, it
is not determined unless a large eccentricity enables us to detect the
periastron advance \citep{seto2001}. For circular binaries, we must
adopt a plausible value of the mass ratio $q$, which we define as the
mass of the lighter member divided by that of the heavier, to infer the
total mass from the chirp mass. If we assume $0.7 \le q \le 1$ motivated
by current observations \citep{tauris_etal2017,ligovirgo2017-3}, the
uncertainty in the total mass is $\approx \SI{2}{per~cent}$. We caution
that $a$ is different from the orbital radius of either component,
$a/(1+q)$ or $qa/(1+q)$.

In fact, the detection of periastron advance improves the situation only
marginally. By following \citet{seto2001}, the error in the eccentricity
is found to be $\Delta{}e\approx\num{3e-3}$ irrespective of the value of
$e$, with assuming that $S_n(3f/2)\approx{}S_n(f)$. The error in the
total mass is determined by the accuracy in determining the frequency of
third orbital harmonics and is at best $\approx \SI{1}{per~cent}$ for
$t_\mathrm{LISA}=\SI{2}{yr}$. This amounts to constraining the mass
ratio to $0.78\lesssim{}q\le1$. We neglect this possible improvement for
simplicity, commenting that a long operation will reduce the error in
the total mass as $t_\mathrm{LISA}^{-3/2}$.

Before concluding this section, we comment on the benefit from a long
operation of \textit{LISA}. Taking $\rho \propto t_\mathrm{LISA}^{1/2}$
shown in equation \eqref{eq:snr} into account, the errors in the phase
(equation \ref{eq:errphase}) and the sky location (equation
\ref{eq:errsky}) decrease as $t_\mathrm{LISA}^{-1/2}$ and
$t_\mathrm{LISA}^{-1}$, respectively, when $t_\mathrm{LISA}$
increases. As we will describe in Section \ref{sec:radio_obs}, these
improvements directly reduce the computational cost required for
detecting radio pulses. Although we adopted the fiducial value of
$t_\mathrm{LISA} = \SI{2}{yr}$, the current nominal plan is \SI{4}{yr}
with a possible extension to \SI{10}{yr} \citep{lisa2017}. Thus, the
statistical errors will be improved compared to those stated in this
paper, and the computational cost may be reduced by a factor of 3--10.

\section{Pulsar survey} \label{sec:radio}

Next, we describe a strategy to detect radio pulses from the shortest
period class of Galactic binary neutron stars utilizing information
obtained by \textit{LISA}. Because \textit{LISA} is planned to be
launched around 2030 \citep{lisa2017}, we aim at detecting the radio
pulses with the Square Kilometer Array (SKA) Phase
2,\footnote{\url{https://astronomers.skatelescope.org/ska/}} which may
complete construction around that time. The radio frequency is denoted
by $\nu$ and the integration time of SKA is denoted by
$t_\mathrm{SKA}$. We mainly focus on observations at popular
\SI{1.4}{\giga\hertz} with the bandwidth $B=\SI{500}{\mega\hertz}$. Our
results can be applied to other facilities by scaling relevant
parameters.

\subsection{Target}

The promising target is a mildly recycled pulsar with the spin period of
$P \sim \SI{30}{\milli\second}$. Indeed, most of the known pulsars in
binary neutron stars are mildly recycled
\citep{tauris_etal2017}. Theoretically, the evolution scenario generally
predicts that one of the neutron stars in these binaries is mildly
recycled and the other is young. The difference in detectability may
come from the beaming fraction, where that of recycled pulsars is
usually considered to be larger than that of young pulsars.  For
example, \citet{levin_etal2013} argue that the former is 0.4--1 compared
to 0.2 of the latter. Our strategy can also find the young pulsar,
whereas it may have a small beaming fraction. In addition, young pulsars
tend to spin down rapidly and could pass the death line before entering
the \textit{LISA} band \citep{tauris_etal2017}.

In this study, we take the fiducial value of the sampling time
$t_\mathrm{samp}$ to be \SI{100}{\micro\second} following plans of
normal pulsar surveys \citep{smits_kslcf2009}. However, we speculate
that $t_\mathrm{samp} \approx \SI{1}{\milli\second}$ may be acceptable
for detecting mildly recycled pulsars. The reason is that the typical
intrinsic pulse width of mildly recycled pulsars is likely to be a few
milliseconds \citep{kramer_xldjwwc1998,lorimer_etal2006} and that the
effective width $W$ must be broader due to dispersion mismatch and
scattering \citep{bhat_ccnl2004}. If we accept this coarse sampling, the
computational cost of data analysis will be reduced substantially, in
particular for a long integration time. Detailed timing observations can
be performed to resolve the pulse profile after the discovery. An
obvious drawback is that the coarse sampling will miss very narrow
pulses. On another front, if the scattering is so severe that the pulses
become wider than their separations (i.e. spin period), observations at
high frequency will be required.

The sensitivity of a search can be estimated by the radiometer equation
\citep{lorimer_kramer}. Following and extending the notation of
\citet{nishino_seto2018}, we write the signal-to-noise ratio
($\mathrm{S/N}$) of a radio source with the flux density $S$ as
\begin{equation}
 \mathrm{S/N} = \gamma \frac{A_\mathrm{eff} S \sqrt{N_\mathrm{p} B
  t_\mathrm{SKA}}}{k_\mathrm{B} T} \sqrt{\frac{P-W}{W}} ,
\end{equation}
where $\gamma$ is a factor of $O(1)$ determined by the detector
configuration, $A_\mathrm{eff}$ is the effective area, $N_\mathrm{p}$ is
the number of polarization modes, and $T$ is the system temperature.
Accordingly, the minimum detectable flux density for a given threshold
of the signal-to-noise ratio, $\mathrm{S/N}_\mathrm{min}$, is given by
\begin{align}
 S_\mathrm{min} & = \SI{e-4}{mJy} \left(
 \frac{\mathrm{S/N}_\mathrm{min}}{5} \right) \left( \frac{\gamma}{1}
 \right)^{-1} \left( \frac{A_\mathrm{eff} /
 T}{\SI{e4}{\square\meter\per\kelvin}} \right)^{-1} \notag \\
 & \times \left( \frac{N_\mathrm{p}}{2} \right)^{-1/2} \left(
 \frac{B}{\SI{500}{\mega\hertz}} \right)^{-1/2} \left(
 \frac{t_\mathrm{SKA}}{\SI{12}{h}} \right)^{-1} \left( \frac{X}{1}
 \right) , \label{eq:smin}
\end{align}
where $X := \sqrt{W/(P-W)}$ becomes unity when $W=0.5P$. The value of
$A_\mathrm{eff} / T$ is normalized aiming at the planned SKA Phase 2
\citep{dewdney_hsl2009}.\footnote{See also URL in the Footnote 1.}

This indicates that 12-h integration of SKA Phase 2 will enable us to
detect pulsars down to the pseudo-luminosity of \SI{0.01}{mJy.kpc^2}
within \SI{10}{kpc} as far as the Doppler smearing is corrected for
appropriately. This sensitivity should be sufficient for finding very
faint pulsars at \SI{1.4}{\giga\hertz}
\citep{kramer_xldjwwc1998,burgay_etal2013}, and non-detection would
indicate the absence of pulses due to the beaming or cessation. The flux
density typically behaves as $\propto \nu^{-1.6}$
\citep{kramer_xldjwwc1998,jankowski_vkbbjk2018}, and thus the search at
high frequency will be a little more challenging. Actual observations
will benefit from the distance estimated to \SI{1}{per~cent} accuracy
with \textit{LISA} to determine the required integration time for a
given pseudo-luminosity.

\subsection{\textit{LISA}-informed radio observation}
\label{sec:radio_obs}

The sky localization of \textit{LISA} ensures covering of the target
pulsar by a single pointing with \SI{15}{\meter} dishes of
SKA. Quantitatively, the size of a pencil beam with the dish diameter
$D_\mathrm{dish}$ is approximately given using the wavelength $\lambda
:= c / \nu$ by \citep{smits_kslcf2009}
\begin{align}
 \Omega_\mathrm{dish} & \approx \frac{\lambda^2}{D_\mathrm{dish}^2} \\
 & = \SI{0.67}{deg^2} \left( \frac{\nu}{\SI{1.4}{\giga\hertz}}
 \right)^{-2} \left( \frac{D_\mathrm{dish}}{\SI{15}{\meter}}
 \right)^{-2} \label{eq:beam}
\end{align}
and is much larger than the localization error of \textit{LISA},
equation \eqref{eq:errsky}. If the entire area of a single-dish beam,
equation \eqref{eq:beam}, is observed by tied-array beams with high
sensitivity in a coherent manner, this information saves the observation
time significantly as we later discuss in Section
\ref{sec:radio_comp}. Accordingly, long integration times could be spent
for observing a single target to achieve high sensitivity as shown in
equation \eqref{eq:smin}. Exceptionally, if we have to look at the
region in the vicinity of the Galactic Center, high frequency of $\sim
\SI{10}{\giga\hertz}$ may be required to overcome severe scattering, and
the single pointing may not be sufficient.

Even if it would be computationally prohibitive to resolve the
single-dish beam completely with an array of sparsely distributed dishes
\citep{smits_kslcf2009,keane_etal2015}, the localization accuracy of
\textit{LISA} is still powerful. The size of a tied-array beam formed by
dishes distributed within $D_\mathrm{array} \sim \SI{1}{\kilo\meter}$ is
given by
\begin{equation}
 \Omega_\mathrm{array} \approx \frac{\lambda^2}{D_\mathrm{array}^2} ,
\end{equation}
where the exact value of $D_\mathrm{array}$ depends on the configuration
and the observing strategy \citep{smits_kslcf2009}. The number of
tied-array beams required to span the localization area of \textit{LISA}
is given by
\begin{align}
 N_\mathrm{array} & \approx \frac{\Delta
 \Omega}{\Omega_\mathrm{array}} \\
 & = 240 \left( \frac{\rho}{200} \right)^{-2} \left(
 \frac{f}{\SI{4}{\milli\hertz}} \right)^2 \left(
 \frac{\nu}{\SI{1.4}{\giga\hertz}} \right)^2
 \left(\frac{D_\mathrm{array}}{\SI{1}{\kilo\meter}} \right)^2 ,
 \label{eq:numbeam}
\end{align}
and this is smaller by a factor of $\Omega_\mathrm{dish} / \Delta \Omega
\approx 20$ than that required to resolve fully a single-dish beam. We
expect that this number of beam forming will be affordable with
computational resources of SKA \citep{keane_etal2015}.

The need to correct for dispersion delay $t_\mathrm{del} ( \nu )$ is not
affected by \textit{LISA} observations unless the Galactic electron
distribution is understood to high accuracy. For completeness, we
estimate the required number of trials for the dispersion measure
(i.e. the column density of free electrons) $\mathrm{DM}$. The step in
trial values is determined by requiring that the delay across the
bandwidth \citep[e.g. section 6.1.1.2 of][]{lorimer_kramer}, which is
approximately given by
\begin{equation}
 t_\mathrm{del} ( \nu - B/2 ) - t_\mathrm{del} ( \nu + B/2 ) \approx
  \frac{e^2}{\pi m_\mathrm{e} c} \mathrm{DM} \frac{B}{\nu^3} ,
\end{equation}
changes by the sampling time. This condition derives the required number
of trials to be
\begin{align}
 N_\mathrm{DM} & \approx \frac{e^2 \mathrm{DM}_\mathrm{max} B}{\pi
 m_\mathrm{e} c \nu^3 t_\mathrm{samp}} \\
 & = \num{15000} \left(
 \frac{\mathrm{DM}_\mathrm{max}}{\SI{1000}{pc.cm^{-3}}} \right) \left(
 \frac{t_\mathrm{samp}}{\SI{100}{\micro\second}} \right)^{-1} \notag \\
 & \times \left( \frac{\nu}{\SI{1.4}{\giga\hertz}} \right)^{-3}
 \left(\frac{B}{\SI{500}{\mega\hertz}} \right) .
\end{align}
This large number of trials is the absolute maximum and would have to be
optimized in realistic data analysis. It is possible that the distance
estimated with \textit{LISA} will weakly constrain the range of
$\mathrm{DM}$.

\begin{figure}
 \includegraphics[width=0.95\linewidth]{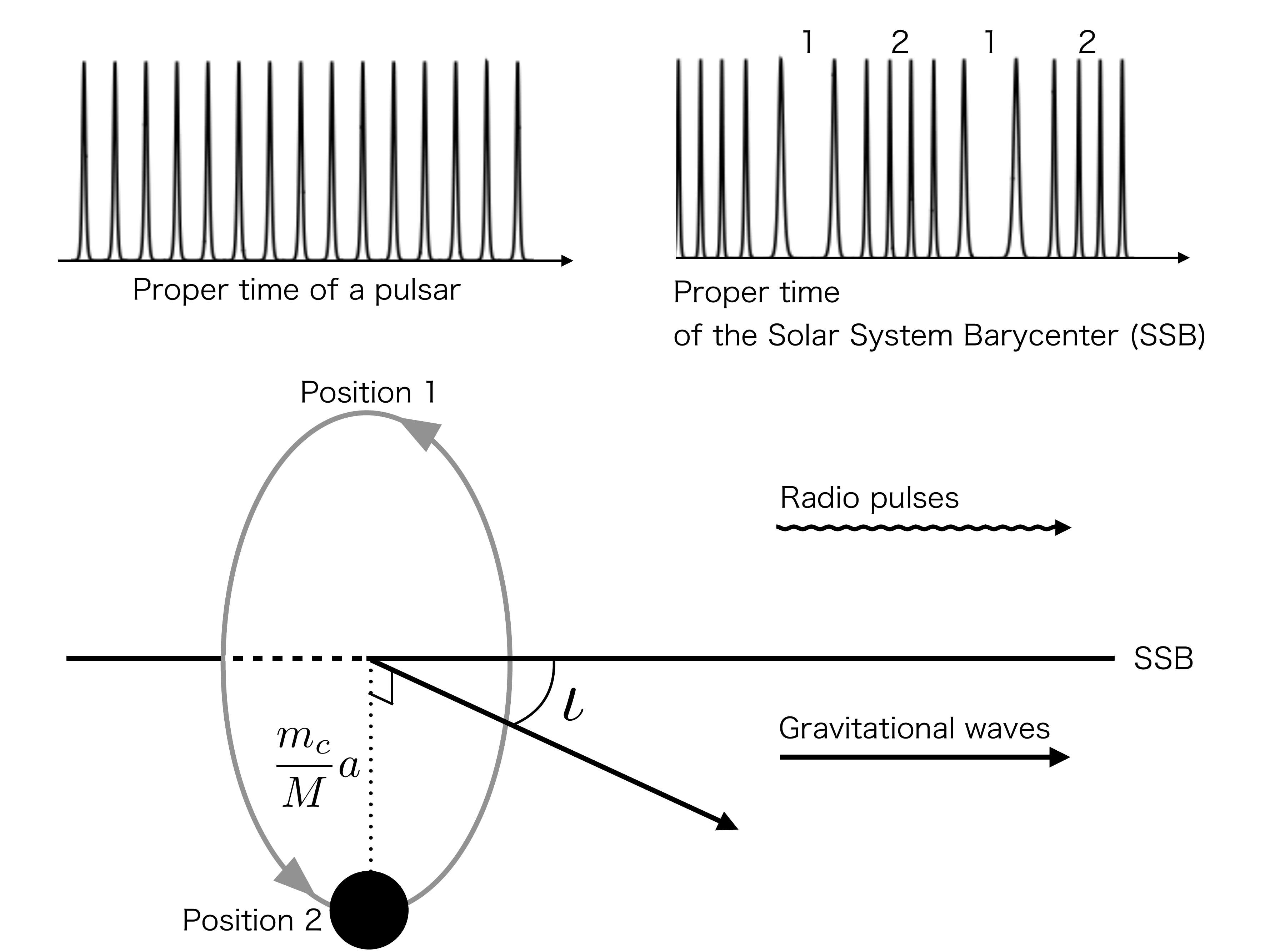} \caption{Schematic
 representation of the orbital modulation of the pulse arrival
 times. Our aim is to recover the original pulse profile on the top left
 from the observed one on the top right by correcting for the modulation
 with information obtained by \textit{LISA}. SSB denotes the Solar
 system barycenter. The inclined arrow on the left denotes the direction
 of orbital angular momentum of the binary. Here, we only draw the
 pulsar without showing the companion in the bottom.}
 \label{fig:schematic}
\end{figure}

\textit{LISA} information will be invaluable for correcting pulse
arrival times for the orbital modulation (see Fig.~\ref{fig:schematic}
for a schematic representation). Because we aim at detecting possibly
faint pulsars in very tight binaries devoting long integration times
that exceed the orbital period, the acceleration search is not
applicable \citep{johnston_kulkarni1991}. While the incoherent
phase-modulation search could be an option
\citep{jouteux_rsjv2002,ransom_ce2003}, the best strategy must be to
integrate coherently the data by correcting the arrival times for the
orbital modulation \cite[see sections 6.2.1 and 6.2.2.1
of][]{lorimer_kramer}. Specifically, if the orbit is circular, we need
to correct them for the amplitude and phase of the modulation. Because
$t_\mathrm{SKA} \ll t_\mathrm{LISA}$, we may safely regard the orbital
frequency as perfectly determined and constant in time.

The modulation amplitude of the pulse arrival time is given in terms of
the projected orbital radius $( m_c / M ) a \sin \iota$ by
\begin{align}
 t_\mathrm{mod} & = \frac{m_c}{M} \frac{a \sin \iota}{c} \\
 & = \SI{0.17}{\second} \left( \frac{m_c / M}{1/2} \right) \left(
 \frac{M}{2.8 M_\odot} \right)^{1/3} \left(
 \frac{f}{\SI{4}{\milli\hertz}} \right)^{-2/3} \left( \frac{\sin
 \iota}{\pi / 4} \right) ,
\end{align}
where $m_c$ is the mass of the companion. Although it is likely that the
recycled pulsar is the heavier component as in the case of J0737$-$3039
\citep{lyne_etal2004} and also suggested by other Galactic binary
neutron stars \citep{tauris_etal2017}, we conservatively think that we
do not know which binary component is detectable as a pulsar \textit{a
priori}. Thus, we need to search pulses for both light and heavy
components.

The uncertainty in $t_\mathrm{mod}$, $\Delta t_\mathrm{mod}$, is given
in terms of the errors in parameters described in Section
\ref{sec:gw}. The errors in the gravitational-wave frequency ($\sim
\num{e-6}$), chirp mass ($< \SI{1}{per~cent}$), and the inclination
($\approx \SI{1}{per~cent}$) are negligible. In fact, the dominant
uncertainty is introduced by the mass ratio, $q$, via $m_c / M$ and $a
\propto M^{1/3}$ irrespective of the error associated with \textit{LISA}
observations. If we assume $0.7 \le q \le 1$, the uncertainty in
$t_\mathrm{mod}$ is $\lesssim \SI{20}{per~cent}$ independent of whether
the pulsar is heavier or lighter. Hereafter, we introduce
$\alpha_\mathrm{amp} := \Delta t_\mathrm{mod} / t_\mathrm{mod}$ with its
fiducial value being $0.2$.

The modulation phase introduces a timing error of $\approx
t_\mathrm{mod} \times \Delta \phi$, where $\Delta \phi \approx 0.015$
for our fiducial parameters, equation \eqref{eq:errphase}. We caution
that, on the basis of the Fisher analysis similar to that performed in
\citet{seto2002}, the accuracy in estimating the gravitational-wave
phase outside the operation period of \textit{LISA} deteriorates so
rapidly that it becomes useless. This means that simultaneous operations
of \textit{LISA} and radio instruments such as SKA are beneficial for
making full use of our multimessenger strategy.

Following consideration of the dispersion measure, we determine the
steps of trial values for the amplitude and phase of the modulation by
requiring that the pulse arrival time changes by the sampling time. The
number of required trials for the modulation amplitude is given by
\begin{align}
 N_\mathrm{amp} & \approx \frac{\Delta
 t_\mathrm{mod}}{t_\mathrm{samp}} \\
 & = 350 \left( \frac{\alpha_\mathrm{amp}}{0.2} \right) \left( \frac{m_c
 / M} {1/2} \right) \left( \frac{M}{2.8 M_\odot} \right)^{1/3} \notag \\
 & \times \left( \frac{f}{\SI{4}{\milli\hertz}} \right)^{-2/3} \left(
 \frac{\sin \iota}{\pi / 4} \right) \left(
 \frac{t_\mathrm{samp}}{\SI{100}{\micro\second}} \right)^{-1} ,
\end{align}
and that for the modulation phase is given by
\begin{align}
 N_\mathrm{phase} & = \frac{t_\mathrm{mod} \Delta
 \phi}{t_\mathrm{samp}} \\
 & = 26 \left( \frac{\rho}{200} \right)^{-1} \left( \frac{m_c / M}{1/2}
 \right) \left( \frac{M}{2.8 M_\odot} \right)^{1/3} \notag \\
 & \times \left( \frac{f}{\SI{4}{\milli\hertz}} \right)^{-2/3} \left(
 \frac{\sin \iota}{\pi / 4} \right) \left(
 \frac{t_\mathrm{samp}}{\SI{100}{\micro\second}} \right)^{-1} .
\end{align}
Additionally, the uncertainty in the eccentricity of $\Delta e \approx
0.003$ could change the arrival time by $t_\mathrm{mod} \Delta e$, and
this will require trials of $N_\mathrm{ecc} \approx 5 ( t_\mathrm{samp}
/ \SI{100}{\micro\second} )$ for our fiducial parameters. We expect that
extension to a slightly eccentric orbit will be straightforward.

We should remark on the spin precession. Its period will be $>
\SI{1}{month} \gg t_\mathrm{SKA}$ for our fiducial parameters
\citep{barker_oconnel1975}, and we expect its effect to be marginal. We
leave the careful assessment as future work, and information from
\textit{LISA} will prove useful even when the precession must be
corrected for.

\subsection{Comparison with an all-sky survey} \label{sec:radio_comp}

To elucidate usefulness of the multimessenger observation with
\textit{LISA}, we compare its performance with an all-sky survey of
radio instruments. To keep the sensitivity fixed, we assume that the
integration time per pointing is the same for both observation
strategies. We also fix the sampling time. A comparison with an all-sky
survey based on the phase-modulation search, with which the sensitivity
is not maximal, is briefly made in the end of this section.

First, the all-sky survey trivially requires a large number of
pointings, and thus the integration time must actually be shortened. If
we do not restrict the search to a limited portion of the sky, e.g. the
Galactic plane, equation \eqref{eq:beam} suggests that $O(\num{e4})$
pointings in terms of a single-dish beam are required to find a limited
number of targets. Even if instruments with a large field-of-view are
available \citep{smits_kslcf2009}, the number of pointings may not be
reduced below $O(\num{e2})$. If we evaluate the number of pointings in
terms of tied-array beams, the all-sky survey requires $O(\num{e8})$
pointings, which is larger by $O(\num{e6})$ compared to the
multimessenger observation with \textit{LISA}, equation
\eqref{eq:numbeam}. However, results of the all-sky survey can be used
for various applications including standard pulsar searches, and thus
this comparison is valid only when we focus on detections of pulsars in
the shortest period class of binary neutron stars.

Next, the coherent integration becomes significantly costly because of
brute-force corrections for the orbital modulation. We have to search
over a large parameter space of not only the modulation amplitude and
phase but also the orbital frequency. By requiring that the phase error
caused by the frequency mismatch accumulated during the integration time
does not induce a timing error larger than the sampling time,
particularly for the harmonic summing to be successful, the step in
trial frequency is given by $\delta f = t_\mathrm{samp} / (2 \pi
t_\mathrm{SKA} t_\mathrm{mod} )$. The number of required trials is
approximately given by $f / \delta f$ and becomes $\sim \num{e6}$ for
our fiducial parameters. The modulation amplitude and phase must also be
searched over without knowledge of the masses and the inclination
angle. On one hand, the number of trial amplitude is increased only by a
factor of $\approx \alpha_\mathrm{amp}^{-1} \approx 5$, because the
uncertainty in the mass ratio is large even if \textit{LISA} information
is available. On the other hand, the number of trial phase is increased
by $\approx 2 \pi / \Delta \phi \approx 400$ for our fiducial
parameters. In total, the number of trails required for coherent
integration per pointing increases by nine orders of magnitude in the
absence of information from \textit{LISA}. This reduces not only the
computational cost but also the false alarm rate so that the threshold
signal-to-noise ratio, $\mathrm{S/N}_\mathrm{min}$, can be lowered.

In reality, the all-sky survey for short-period binaries will likely be
conducted with incoherent but efficient phase-modulation searches, and
thus it would be informative to compare performance of our strategy with
theirs. The number of pointings is reduced significantly by
\textit{LISA} as described above. If the trials of corrections for the
orbital modulation can be performed in the Fourier space for our
coherent search \citep{ransom_ghmcel2001}, the computational cost for
each pointing may be dominated by the Fourier transformation of time
series data. If this is realized, the costs will be similar for both
methods as long as the integration time and the sampling time are
fixed. Otherwise, the computational cost of our strategy could be higher
by a factor of $N_\mathrm{amp} N_\mathrm{phase} N_\mathrm{ecc} \approx
O(10^4)$ than the phase-modulation search. This very approximately
cancels the gain due to the sky localization. Still, the smaller number
of required pointings allows our strategy to devote longer integration
times at each sky area. In both cases, the sensitivity of the
phase-modulation search is limited to, e.g. $\approx
30$--\SI{40}{per~cent} of the coherent search for the parameters of
fig.~6 of \citet{ransom_ce2003}.  Thus, the detectable volume will be
reduced accordingly for a given integration time. We feel that a further
comparison of the performance is difficult because of the different
sensitivity. A useful comparison may be made by conducting,
e.g. synthetic surveys with simulated populations of short-period binary
pulsars, which is beyond the scope of this study.

\section{Summary} \label{sec:summary}

We have studied a multimessenger strategy to detect pulsars in the
shortest period class of Galactic binary neutron stars with
\textit{LISA} and radio instruments such as SKA. These binaries will be
observed by \textit{LISA} as quasi-monochromatic gravitational-wave
sources with a large signal-to-noise ratio of $>100$. A single pointing
in terms of a single dish of SKA will be sufficient to entirely cover
the localization area of \textit{LISA}, and accordingly long integration
times may become available to improve the sensitivity. While this long
integration time prohibits the acceleration search, the Doppler smearing
caused by the rapid orbital motion can be corrected efficiently by using
orbital frequency and binary parameters derived with \textit{LISA}.

The multimessenger observation, or follow-up surveys of \textit{LISA}
detections, will reduce the numbers of required pointings and trials of
corrections for orbital modulations by a factor of \num{e2}--\num{e6}
and \num{e9}, respectively, compared to an all-sky survey of the
shortest period class of binary neutron stars for given sensitivity. It
is highly likely that realistic situations limit the integration time
per pointing for an all-sky survey significantly and that we will miss
faint and distant pulsars. The massive computational cost of the survey
without \textit{LISA} information can be mitigated by incoherent search
techniques but only at the expense of the sensitivity. Finally, if
observations of \textit{LISA} and SKA do not occur simultaneously, we
will lose information of the phase so that the efficiency of the
multimessenger strategy will be reduced by two orders of
magnitude. Although the improvement is still significant, simultaneous
observations are preferable.

\section*{Acknowledgements}

We thank Shota Kisaka for valuable discussions. This work is supported
by Japanese Society for the Promotion of Science (JSPS) KAKENHI Grant
Numbers JP15K05075, JP16H06342, JP17H01131, JP17H06358, JP18H04595, and
JP18H05236.

%%%%%%%%%%%%%%%%%%%%%%%%%%%%%%%%%%%%%%%%%%%%%%%%%%



\bibliographystyle{mnras}
%\bibliography{paper} % if your bibtex file is called example.bib

\begin{thebibliography}{}
\makeatletter
\relax
\def\mn@urlcharsother{\let\do\@makeother \do\$\do\&\do\#\do\^\do\_\do\%\do\~}
\def\mn@doi{\begingroup\mn@urlcharsother \@ifnextchar [ {\mn@doi@}
  {\mn@doi@[]}}
\def\mn@doi@[#1]#2{\def\@tempa{#1}\ifx\@tempa\@empty \href
  {http://dx.doi.org/#2} {doi:#2}\else \href {http://dx.doi.org/#2} {#1}\fi
  \endgroup}
\def\mn@eprint#1#2{\mn@eprint@#1:#2::\@nil}
\def\mn@eprint@arXiv#1{\href {http://arxiv.org/abs/#1} {{\tt arXiv:#1}}}
\def\mn@eprint@dblp#1{\href {http://dblp.uni-trier.de/rec/bibtex/#1.xml}
  {dblp:#1}}
\def\mn@eprint@#1:#2:#3:#4\@nil{\def\@tempa {#1}\def\@tempb {#2}\def\@tempc
  {#3}\ifx \@tempc \@empty \let \@tempc \@tempb \let \@tempb \@tempa \fi \ifx
  \@tempb \@empty \def\@tempb {arXiv}\fi \@ifundefined
  {mn@eprint@\@tempb}{\@tempb:\@tempc}{\expandafter \expandafter \csname
  mn@eprint@\@tempb\endcsname \expandafter{\@tempc}}}

\bibitem[\protect\citeauthoryear{{Abbott} et~al.,}{{Abbott}
  et~al.}{2017}]{ligovirgo2017-3}
{Abbott} B.~P.,  et~al., 2017, \mn@doi [\prl] {10.1103/PhysRevLett.119.161101},
  119, 161101

\bibitem[\protect\citeauthoryear{{Abbott} et~al.,}{{Abbott}
  et~al.}{2018}]{ligovirgo2018-4}
{Abbott} B.~P.,  et~al., 2018, arXiv:1811.00364

\bibitem[\protect\citeauthoryear{{Amaro-Seoane} et~al.,}{{Amaro-Seoane}
  et~al.}{2017}]{lisa2017}
{Amaro-Seoane} P.,  et~al., 2017, arXiv:1702.00786

\bibitem[\protect\citeauthoryear{Andersen \& Ransom}{Andersen \&
  Ransom}{2018}]{andersen_ransom2018}
Andersen B.~C.,  Ransom S.~M.,  2018, \mn@doi [\apj]
  {10.3847/2041-8213/aad59f}, 863, L13

\bibitem[\protect\citeauthoryear{{Armano} et~al.,}{{Armano}
  et~al.}{2016}]{lisapf2016}
{Armano} M.,  et~al., 2016, \mn@doi [\prl] {10.1103/PhysRevLett.116.231101},
  116, 231101

\bibitem[\protect\citeauthoryear{{Armano} et~al.,}{{Armano}
  et~al.}{2018}]{lisapf2018}
{Armano} M.,  et~al., 2018, \mn@doi [\prl] {10.1103/PhysRevLett.120.061101},
  120, 061101

\bibitem[\protect\citeauthoryear{Bagchi, Lorimer  \& Wolfe}{Bagchi
  et~al.}{2013}]{bagchi_lw2013}
Bagchi M.,  Lorimer D.~R.,   Wolfe S.,  2013, \mn@doi [\mnras]
  {10.1093/mnras/stt559}, 432, 1303

\bibitem[\protect\citeauthoryear{Barker \& O'Connell}{Barker \&
  O'Connell}{1975}]{barker_oconnel1975}
Barker B.~M.,  O'Connell R.~F.,  1975, \mn@doi [\prd]
  {10.1103/PhysRevD.12.329}, 12, 329

\bibitem[\protect\citeauthoryear{Bhat, Cordes, Camilo, Nice  \& Lorimer}{Bhat
  et~al.}{2004}]{bhat_ccnl2004}
Bhat N. D.~R.,  Cordes J.~M.,  Camilo F.,  Nice D.~J.,   Lorimer D.~R.,  2004,
  \mn@doi [\apj] {10.1086/382680}, 605, 759

\bibitem[\protect\citeauthoryear{Burgay et~al.,}{Burgay
  et~al.}{2013}]{burgay_etal2013}
Burgay M.,  et~al., 2013, \mn@doi [\mnras] {10.1093/mnras/stt721}, 433, 259

\bibitem[\protect\citeauthoryear{Cameron et~al.,}{Cameron
  et~al.}{2018}]{cameron_etal2018}
Cameron A.~D.,  et~al., 2018, \mn@doi [\mnras] {10.1093/mnrasl/sly003}, 475,
  L57

\bibitem[\protect\citeauthoryear{Cutler}{Cutler}{1998}]{cutler1998}
Cutler C.,  1998, \mn@doi [\prd] {10.1103/PhysRevD.57.7089}, 57, 7089

\bibitem[\protect\citeauthoryear{Dewdney, Peter, Schlizzi  \& Lazio}{Dewdney
  et~al.}{2009}]{dewdney_hsl2009}
Dewdney P.~E.,  Peter H.~J.,  Schlizzi R.~T.,   Lazio T. J. L.~W.,  2009,
  \mn@doi [IEEE Proceedings] {10.1109/JPROC.2009.2021005}, 97, 1482

\bibitem[\protect\citeauthoryear{Farmer \& Phinney}{Farmer \&
  Phinney}{2003}]{farmer_phinney2003}
Farmer A.,  Phinney E.~S.,  2003, \mn@doi [\mnras]
  {10.1111/j.1365-2966.2003.07176.x}, 346, 1197

\bibitem[\protect\citeauthoryear{{Faulkner} et~al.,}{{Faulkner}
  et~al.}{2004}]{faulkner_etal2004}
{Faulkner} A.~J.,  et~al., 2004, \mn@doi [\mnras]
  {10.1111/j.1365-2966.2004.08310.x}, 355, 147

\bibitem[\protect\citeauthoryear{Hulse \& Taylor}{Hulse \&
  Taylor}{1975}]{hulse_taylor1975}
Hulse R.~A.,  Taylor J.~H.,  1975, \mn@doi [\apj] {10.1086/181708}, 195, L51

\bibitem[\protect\citeauthoryear{Jacoby, Cameron, Jenet, Anderson, Murty  \&
  Kulkarni}{Jacoby et~al.}{2006}]{jacoby_cjamk2006}
Jacoby B.~A.,  Cameron P.~B.,  Jenet F.~A.,  Anderson S.~B.,  Murty R.~N.,
  Kulkarni S.~R.,  2006, \mn@doi [\apj] {10.1086/505742}, 644, L113

\bibitem[\protect\citeauthoryear{Jankowski, {van Straten}, Keane, Bailes, Barr,
  Johnston  \& Kerr}{Jankowski et~al.}{2018}]{jankowski_vkbbjk2018}
Jankowski F.,  {van Straten} W.,  Keane E.~F.,  Bailes M.,  Barr E.,  Johnston
  S.,   Kerr M.,  2018, \mn@doi [\mnras] {10.1093/mnras/stx2476}, 473, 4436

\bibitem[\protect\citeauthoryear{Johnston \& Kulkarni}{Johnston \&
  Kulkarni}{1991}]{johnston_kulkarni1991}
Johnston H.~M.,  Kulkarni S.~R.,  1991, \mn@doi [\apj] {10.1086/169715}, 368,
  504

\bibitem[\protect\citeauthoryear{Jouteux, Ramachandran, Stappers, Jonker  \&
  {van der Klis}}{Jouteux et~al.}{2002}]{jouteux_rsjv2002}
Jouteux S.,  Ramachandran R.,  Stappers B.~W.,  Jonker P.,   {van der Klis} M.,
   2002, \mn@doi [\aap] {10.1051/0004-6361:20020052}, 384, 532

\bibitem[\protect\citeauthoryear{{Keane} et~al.,}{{Keane}
  et~al.}{2015}]{keane_etal2015}
{Keane} E.~F.,  et~al., 2015, Advancing Astrophysics with the Square Kilometre
  Array (AASKA14), p.~40

\bibitem[\protect\citeauthoryear{Kramer, Xilouris, Lorimer, Doroshenko,
  Jessner, Wielebinski, Wolszczan  \& Camilo}{Kramer
  et~al.}{1998}]{kramer_xldjwwc1998}
Kramer M.,  Xilouris K.~M.,  Lorimer D.~R.,  Doroshenko O.,  Jessner A.,
  Wielebinski R.,  Wolszczan A.,   Camilo F.,  1998, \mn@doi [\apj]
  {10.1086/305790}, 501, 270

\bibitem[\protect\citeauthoryear{Kyutoku \& Seto}{Kyutoku \&
  Seto}{2016}]{kyutoku_seto2016}
Kyutoku K.,  Seto N.,  2016, \mn@doi [\mnras] {10.1093/mnras/stw1767}, 462,
  2177

\bibitem[\protect\citeauthoryear{Levin et~al.,}{Levin
  et~al.}{2013}]{levin_etal2013}
Levin L.,  et~al., 2013, \mn@doi [\mnras] {10.1093/mnras/stt1103}, 434, 1387

\bibitem[\protect\citeauthoryear{Lorimer \& Kramer}{Lorimer \&
  Kramer}{2004}]{lorimer_kramer}
Lorimer D.~R.,  Kramer M.,  2004, Handbook of pulsar astronomy.
Cambridge University Press

\bibitem[\protect\citeauthoryear{{Lorimer} et~al.,}{{Lorimer}
  et~al.}{2006}]{lorimer_etal2006}
{Lorimer} D.~R.,  et~al., 2006, \mn@doi [\mnras]
  {10.1111/j.1365-2966.2006.10887.x}, 372, 777

\bibitem[\protect\citeauthoryear{Lyne et~al.,}{Lyne
  et~al.}{2004}]{lyne_etal2004}
Lyne A.~G.,  et~al., 2004, \mn@doi [Science] {10.1126/science.1094645}, 303,
  1153

\bibitem[\protect\citeauthoryear{Nelemans, Yungelson  \& {Portegies
  Zwart}}{Nelemans et~al.}{2004}]{nelemans_yp2004}
Nelemans G.,  Yungelson L.~R.,   {Portegies Zwart} S.~F.,  2004, \mn@doi
  [\mnras] {10.1111/j.1365-2966.2004.07479.x}, 349, 181

\bibitem[\protect\citeauthoryear{Nishino \& Seto}{Nishino \&
  Seto}{2018}]{nishino_seto2018}
Nishino Y.,  Seto N.,  2018, \mn@doi [\apj] {10.3847/2041-8213/aad33d}, 862,
  L21

\bibitem[\protect\citeauthoryear{Peters}{Peters}{1964}]{peters1964}
Peters P.~C.,  1964, \mn@doi [Physical Review] {10.1103/PhysRev.136.B1224},
  136, B1224

\bibitem[\protect\citeauthoryear{Prince, Anderson, Kulkarni  \&
  Wolszczan}{Prince et~al.}{1991}]{prince_akw1991}
Prince T.~A.,  Anderson S.~B.,  Kulkarni S.~R.,   Wolszczan A.,  1991, \mn@doi
  [\apj] {10.1086/186067}, 374, L41

\bibitem[\protect\citeauthoryear{Ransom, Greenhill, Herrnstein, Manchester,
  Camilo, Eikenberry  \& Lyne}{Ransom et~al.}{2001}]{ransom_ghmcel2001}
Ransom S.~M.,  Greenhill L.~J.,  Herrnstein J.~R.,  Manchester R.~N.,  Camilo
  F.,  Eikenberry S.~S.,   Lyne A.~G.,  2001, \mn@doi [\apj] {10.1086/318062},
  546, L25

\bibitem[\protect\citeauthoryear{Ransom, Cordes  \& Eikenberry}{Ransom
  et~al.}{2003}]{ransom_ce2003}
Ransom S.~M.,  Cordes J.~M.,   Eikenberry S.~S.,  2003, \mn@doi [\apj]
  {10.1086/374806}, 589, 911

\bibitem[\protect\citeauthoryear{Robson, Cornish  \& Liu}{Robson
  et~al.}{2018}]{robson_cl2018}
Robson T.,  Cornish N.,   Liu C.,  2018, arXiv:1803.01944

\bibitem[\protect\citeauthoryear{Seto}{Seto}{2001}]{seto2001}
Seto N.,  2001, \mn@doi [\prl] {10.1103/PhysRevLett.87.251101}, 87, 251101

\bibitem[\protect\citeauthoryear{Seto}{Seto}{2002}]{seto2002}
Seto N.,  2002, \mn@doi [\mnras] {10.1046/j.1365-8711.2002.05432.x}, 333, 469

\bibitem[\protect\citeauthoryear{Smits, Kramer, Stappers, Lorimer, Cordes  \&
  Faulkner}{Smits et~al.}{2009}]{smits_kslcf2009}
Smits R.,  Kramer M.,  Stappers B.,  Lorimer D.~R.,  Cordes J.,   Faulkner A.,
  2009, \mn@doi [\aap] {10.1051/0004-6361:200810383}, 493, 1161

\bibitem[\protect\citeauthoryear{Stovall et~al.,}{Stovall
  et~al.}{2018}]{stovall_etal2018}
Stovall K.,  et~al., 2018, \mn@doi [\apj] {10.3847/2041-8213/aaad06}, 854, L22

\bibitem[\protect\citeauthoryear{Takahashi \& Seto}{Takahashi \&
  Seto}{2002}]{takahashi_seto2002}
Takahashi R.,  Seto N.,  2002, \mn@doi [\apj] {10.1086/341483}, 575, 1030

\bibitem[\protect\citeauthoryear{Tauris et~al.,}{Tauris
  et~al.}{2017}]{tauris_etal2017}
Tauris T.~M.,  et~al., 2017, \mn@doi [\apj] {10.3847/1538-4357/aa7e89}, 846,
  170

\bibitem[\protect\citeauthoryear{Weisberg \& Huang}{Weisberg \&
  Huang}{2016}]{weisberg_huang2016}
Weisberg J.~M.,  Huang Y.,  2016, \mn@doi [\apj] {10.3847/0004-637X/829/1/55},
  829, 55

\bibitem[\protect\citeauthoryear{Will}{Will}{2014}]{will2014}
Will C.~M.,  2014, \mn@doi [Living Reviews in Relativity]
  {10.12942/lrr-2014-4}, 17, 4

\bibitem[\protect\citeauthoryear{Yao, Manchester  \& Wang}{Yao
  et~al.}{2017}]{yao_mw2017}
Yao J.~M.,  Manchester R.~N.,   Wang N.,  2017, \mn@doi [\apj]
  {10.3847/1538-4357/835/1/29}, 835, 29

\bibitem[\protect\citeauthoryear{Yunes, Yagi  \& Pretorius}{Yunes
  et~al.}{2016}]{yunes_yp2016}
Yunes N.,  Yagi K.,   Pretorius F.,  2016, \mn@doi [\prd]
  {10.1103/PhysRevD.94.084002}, 94, 084002

\makeatother
\end{thebibliography}


%%%%%%%%%%%%%%%%%%%%%%%%%%%%%%%%%%%%%%%%%%%%%%%%%%

%%%%%%%%%%%%%%%%% APPENDICES %%%%%%%%%%%%%%%%%%%%%

%\appendix

%%%%%%%%%%%%%%%%%%%%%%%%%%%%%%%%%%%%%%%%%%%%%%%%%%

% Don't change these lines
\bsp	% typesetting comment
\label{lastpage}
\end{document}